\documentclass[12pt]{article}

\usepackage{amssymb,amsmath,amsthm,amscd,latexsym}
\usepackage{mathrsfs}
\usepackage{mathrsfs}
\usepackage{amsfonts}
\usepackage{amsmath}
\usepackage{amssymb}
\usepackage{amscd}
\usepackage{mathrsfs}
\usepackage{amssymb}
\usepackage{amsmath}
\usepackage{amsthm}
\usepackage{latexsym}
\usepackage{indentfirst}
\usepackage{enumitem}
\usepackage{anysize}
\usepackage{bbm}
\usepackage{arydshln}
\usepackage[pdftex]{graphicx}

\renewcommand{\paragraph}{\roman{paragraph}}
\input cyracc.def

\newcommand{\F}{\mathbb{F}}

\newtheorem{theorem}{Theorem}
\newtheorem{cor}{Corollary}

\newtheorem{lemma}{Lemma}
\newtheorem{prop}{Proposition}

\theoremstyle{definition}

\begin{document}
\title{\bf How many weights can a linear code have?
\thanks{This research is supported by National Natural Science Foundation of China (61672036), Excellent Youth Foundation of Natural Science Foundation of Anhui Province (1808085J20),
Technology Foundation for Selected Overseas Chinese Scholar, Ministry of Personnel of China (05015133) and
Key projects of support program for outstanding young talents in Colleges and Universities (gxyqZD2016008).}}

\author{
\small{Minjia Shi$^{1,2}$, Hongwei Zhu$^2$, Patrick Sol\'e$^3$, and G\'erard D. Cohen$^4$}\\ 
\and \small{${}^1$Key Laboratory of Intelligent Computing \& Signal Processing,}\\
 \small{Ministry of Education, Anhui University No. 3 Feixi Road,}\\
  \small{Hefei Anhui Province 230039, P. R. China;}\\
\small{${}^2$School of Mathematical Sciences, Anhui University, Hefei, 230601, China}\\
[-0.8ex]
\small{${}^3$CNRS/LAGA, University of Paris 8, 2 rue de la Libert\'e, 93 526 Saint-Denis, France}\\
\small{${}^4$ TelecomParisTech, 46 rue Barrault, 75 013 Paris}}
\date{}
\maketitle
\begin{abstract}  We study the combinatorial function $L(k,q),$ the maximum number of nonzero weights a linear code of dimension $k$ over $\F_q$ can have.
We determine it completely for $q=2,$ and for $k=2,$ and provide upper and lower bounds in the general case when both $k$ and $q$ are  $\ge 3.$
A refinement $L(n,k,q),$ as well as nonlinear analogues $N(M,q)$ and $N(n,M,q),$ are also introduced and studied.
\end{abstract}


\section{Introduction}
There are several problems in extremal combinatorics on distances in codes. For instance, the famous paper \cite{D} derives an upper bound on the size of a code $C$ over $\F_q$ with exactly $s$ distinct distances:
\begin{equation}
|C|\le \sum_{j=0}^s {n \choose j} (q-1)^j.
\end{equation}
In the same spirit, other authors have given upper bounds on the size of codes with one or several forbidden distances \cite{EFIN}.

 In this note, we tackle a related but distinctly different problem:
how many distinct weights can a linear code of given dimension over a given finite field have ? In other words, we study the combinatorial function $L(k,q),$ the maximum number
of nonzero weights a code of dimension $k$ over $\F_q$ may have. While an upper bound is easy to prove (Proposition \ref{prop1}),
its tightness is nontrivial\footnote{ After submission of this article, a proof was found in \cite{proof}.} and we only manage to establish it in some special cases like $k=2$ or $q=2$
(Cf. Theorem 1 and Theorem 2). Numerical experiments with very long random codes suggest it is tight for all $k$'s and $q$'s. We leave the question as an open problem.
We can also study the more refined function $L(n,k,q),$ the maximum number of nonzero weights an $[n,k]_q$ code may have. This latter function is related to both $L(k,q)$ and
equation $(1)$ above.
The nonlinear counterpart of $L(k,q)$ denoted by  $N(M,q),$ can be determined explicitly (Theorem \ref{tight2}). The nonlinear counterpart of $L(n,k,q)$ denoted by  $N(n,M,q),$
can also be studied. The rate of convergence of $N(n,M,q)$ towards $N(M,q)$ requires perfect difference sets \cite{BJL} and primes in short intervals \cite{BHP} for its careful study.

The material is organized as follows. Section 2 collects the necessary notations and definitions. Section 3 studies upper bounds in the linear code case.
Section 4 derives lower bounds in that situation. Section 5 introduces and investigates the function $L(n,k,q).$
Section 6 tackles the nonlinear analogues of $L(k,q)$ and $L(n,k,q),$ denoted by $N(M,q),$ and $N(n,M,q),$ respectively. Section 7 concludes the article.
An appendix collects some numerical values, which comfort the Conjecture that Proposition \ref{prop1} is tight.
\section{Definitions and notation}
Let $q$ be a prime power, and $\F_q$ denote the finite field of order $q.$
By a {\bf code} of length $n$ over $\F_q,$ we shall mean a proper subset of $\F_q^n.$ This code is {\bf linear} if it is a  $\F_q$-vector subspace of $\F_q^n.$
 The {\bf dimension} of a code, denoted by $k$, is equal to its dimension as a vector space. The parameters of such a code are written compactly as $[n,k]_q.$
The {\bf Hamming weight}  of $x\in \F_q^n,$ denoted by $w(x),$ is the number of indices $i$ where $x_i \neq 0.$
The {\bf Hamming distance}  between $x\in \F_q^n,$ and $y \in \F_q^n,$ denoted by $d(x,y),$ is defined by $d(x,y)=w(x-y).$
For a given prime power $q$ and given values of $k$, let {\bf $L(k,q)$} denote the largest possible number of nonzero weights a $q$-ary code can have.
If $C(n)$ is a family of codes of parameters $[n, k_n]_q$, the {\bf rate} $R$ is defined as $$R=\limsup\limits_{n \rightarrow \infty}\frac{k_n}{n}.$$
Recall that the $q$-ary {\bf entropy function} $H_q(.)$ is defined for $0<y< 1,$ by $$ H_q(y)=y\log_q(q-1)-y\log_q(y)-(1-y)\log_q(1-y).$$
\section{Upper bounds}
The following monotonicity properties of $L(k,q)$ are given without proof.
{\prop For all nonegative integers $k,m$ and all prime powers $q$ we have:
\begin{eqnarray*}
L(k,q)&\le & L(k+1,q),\\
L(k,q)&\le & L(k,q^m).
\end{eqnarray*}
}
The next result is trivial but crucial.
{\prop \label{prop1} For all prime powers $q,$ and all integers $k\ge 1,$ we have
$$L(k,q)\le \frac{q^k-1}{q-1}.$$
}\vspace{-0.8cm}
\begin{proof}
The total numbers of nonzero codewords of a code of dimension $k$ over $\F_q$ is $q^k-1,$ and all the nonzero multiples of a given codeword share the same weight.
\end{proof}

This bound is met with equality if $q=2.$
{\theorem \label{tight} For all integers $k\ge 1,$ we have
$$L(k,2)= {2^k-1}.$$}\vspace{-1.0cm}
\begin{proof}
Denote by $G_k$ the generator matrix of an $[n,k]_q$ with $L(k,2)$ weights $w_1<w_2<\dots<w_{L(k,2)}.$
Define $H_{k+1}$ a matrix obtained from $G_k$ by adding a $k$ by $t$ block of zeros, and by  $G_{k+1}$ the matrix obtained by $H_{k+1}$ by adding an
a row with first $n$ coordinates zero and last $t$ coordinates $=1.$
The code spanned by the rows of $G_{k+1}$ has all these weights plus the $ L(k,2)+1$ new weights $t<t+w_1<\cdots<t+w_{L(k,2)}.$ The two sets of weights will have void intersection if
$w_{L(k,2)}<t.$  This makes $2L(k,2)+1$ weights altogether. Note that the rank of $G_{k+1}$ is $k+1.$
Thus we have proved that  $L(k+1,2)\ge 2L(k,2)+1,$ which implies by induction, starting from $L(1,2)=1,$ the lower bound $L(k,2)\ge{2^k-1}.$ The result follows.\end{proof}
{\bf Remark:} We are now ready to given an alternative proof of Theorem 1. we can exhibit a linear code $C$ with dimension $k$ over $\F_2$ with $2^k-1$ nonzero weights. Let the generator matrix of $C$ be
\[
\begin{array}{c@{\hspace{-5pt}}l}
\left(\begin{array}{c;{1pt/1pt}cc;{2pt/2pt}cccc;{2pt/2pt}cccccccc;{2pt/2pt}c;{2pt/2pt}ccccc}
  1      & 1      & 1      & 1 & 1 & 1 & 1   &1&1&1&1&1&1&1&1& \cdots & 1 & 1 & \cdots & 1 & 1\\
    1      & 1      & 1      & 1 & 1 & 1 & 1   &1&1&1&1&1&1&1&1& \cdots & 0 & 0 & \cdots & 0 & 0\\
    \vdots & \vdots & \vdots & \vdots & \vdots &\vdots&\vdots&\vdots&\vdots&\vdots&\vdots&\vdots&\vdots& \vdots & \vdots & \vdots & \vdots & \vdots & \vdots & \vdots & \vdots\\
    1      & 1      & 1      & 1 & 1 & 1 & 1   &1&1&1&1&1&1&1&1& \cdots & 0 & 0 & \cdots & 0 & 0\\
    1      & 1      & 1      & 1 & 1 & 1 & 1   &0&0&0&0&0&0&0&0& \cdots & 0 & 0 & \cdots & 0 & 0\\
    1      & 1      & 1      & 0 & 0 & 0 & 0   &0&0&0&0&0&0&0&0& \cdots & 0 & 0 & \cdots & 0 & 0\\
    1      & 0      & 0      & 0 & 0 & 0 & 0   &0&0&0&0&0&0&0&0& \cdots & 0 & 0 & \cdots & 0 & 0\\
\end{array}\right),
&\\[-4pt]
\begin{array}{cccccccccc}
\hspace*{-0.5cm}\underbrace{\rule{0.05cm}{0mm}}_{a_1}&
\hspace*{-0.3cm}\underbrace{\rule{9mm}{0mm}}_{a_2} &
\hspace*{-0.15cm}\underbrace{\rule{21mm}{0mm}}_{a_3} &
\hspace*{-0.1cm}\underbrace{\rule{4.2cm}{0mm}}_{a_4} &
\hspace*{0.6cm}\underbrace{\rule{3cm}{0mm}}_{a_k}\end{array} &
\end{array}
\]
where $a_1=1,$ $a_2=2$, $a_3=2^2$, $a_4=2^3$, $\ldots$, $a_k=2^{k-1}.$
Since $a_{j_1}+a_{j_2}+\ldots+a_{j_t}= \underbrace{{(\ldots010\ldots010\ldots010\ldots)}_2}_k$ in base $2,$ and the coordinates of $1'$s are $j_1,j_2,\ldots,j_t$, respectively,
it can be seen that the Hamming weight of $uG$ is equal to the integer whose expansion in base $2$ is $u.$
Thus, we obtain all integers of $k$ bits as possible weights that is the set
$\{1,2,3,\dots,2^k-1\}$
of cardinality $2^k-1$ in all.

The bound in proposition \ref{prop1} is also tight when $k=2.$
{\theorem For all prime powers $q,$  we have
$L(2,q)= q+1.$
}
\begin{proof}
Let $\{u,v\}$ be a basis of a code $C$ candidate to have $q+1$ weights. Denote by $S,\,T$ the supports of $u,v$ respectively. Let $|S\setminus T|=a,$ $|T\setminus S|=b.$
On the intersection $S\bigcap T$ assume $v$ is the all-one vector. Denote by $\omega$ a primitive root of $\F_q.$ Assume $|S\bigcap T|={q \choose 2}$ and that $u$ restricted to $S\bigcap T$ is
$$(1,\omega, \omega, \omega^2,\omega^2,\omega^2,\dots,\omega^{q-2},\dots,\omega^{q-2}) $$
where $\omega^i$ occurs $i+1$ times. With these conventions, we see that the weights of $C$ are
\begin{itemize}
\item $w(u)=a+{q \choose 2},$
\item $w(v)=b+{q \choose 2},$
\item $w(u-xv)=a+b+ {q \choose 2}-i$ if $x=\omega^{i-1}$ for $i=1,2,\dots,q-1.$
\end{itemize}
Assume $a<b.$ The above weights will be pairwise different if $a+b+{q \choose 2}-(q-1)>b+{q \choose 2},$ that is if $a\ge q.$
Thus, under these conditions, $C$ counts $2+q-1=q+1$ nonzero weights.
\end{proof}

{\bf Remark:} The shortest $[n,2]_q$ code with $L(2,q)$ nonzero weights obtained by this construction has $n={q \choose 2}+2q+1.$

\section{Lower bounds}
The easiest lower bound is

{\prop For all prime powers $q,$ and all integers $k\ge 1,$ we have
$L(k,q)\ge k.$
}
\begin{proof}
Consider the code $\F_q^k,$ of length and dimension $k.$
\end{proof}

This can be improved to a bound that is exponential in $k.$
{\prop \label{expo} For all prime powers $q,$ and all integers $k\ge 1,$ we have
$$L(k+1,q)\ge 2L(k,q)+1.$$ In particular, for all integers $k\ge 2,$ we have $$L(k,q)\ge 2^{k-2}q+2^{k-2}+1.$$
} \vspace{-0.8cm}
\begin{proof}
Same argument as in the first proof of Theorem \ref{tight}. The second assertion follows by iterating this bound starting from $L(2,q)=q+1.$
\end{proof}
An asymptotic version of the preceding results is as follows. Define

$$ \lambda(q)=\limsup_{n \rightarrow \infty} \frac{1}{k}\log_q(L(k,q)) .$$

{\theorem For all prime powers $q$ we have $$\log_q 2 \le \lambda(q) \le 1.$$ In particular $\lambda(2)=1.$
}
\begin{proof}
The first inequality comes from Proposition \ref{expo}. The second one comes Proposition \ref{prop1}.
\end{proof}

{\bf Remark:} Since we conjecture that the bound of Proposition \ref{prop1} is tight, it is natural to conjecture that $\lambda(q)=1$ for all prime powers $q.$

\section{Refinements and asymptotics}

A more complex function is $L(n,k,q)$ the largest number of nonzero weights an $[n,k]_q$-code can have. This function is related to $L(k,q)$ in several ways.
The following monotonicity properties of $L(n,k,q)$ are given without proof.
{\prop For all nonegative integers $k,m$ and all prime powers $q$ we have:
\begin{eqnarray*}
L(n,k,q)&\le & L(n,k+1,q),\\
L(n,k,q)&\le & L(n,k,q^m).
\end{eqnarray*}
}
The three following lemmas are useful for the proof of Theorem \ref{thm4}.
{\lemma \label{1} For all prime powers $q,$ and all nonnegative integers $n,k$ we have
$L(n,k, q)\le L(k,q).$
}
\begin{proof}
Immediate from the definitions.
\end{proof}

The new function is also monotone in $n.$
{\lemma \label{2} For all prime powers $q,$ and all nonnegative integers $n,k$ we have
$L(n,k, q)\le L(n+1, k,q).$
}
\begin{proof}
If $C$ is an $[n,k]_q$ code with $L(n,k, q)$ nonzero weights, then $C$ extended by a constant zero coordinate is an $[n+1,k]_q$-code with the same number of nonzero weights.
\end{proof}

{\lemma \label{3} For all prime powers $q,$ and all nonnegative integers $n,k$ we have
$L(n,k, q)\le n.$
}
\begin{proof}
Note that, by definition of the Hamming weight, a code of length $n$ can have at most $n$ distinct weights.
\end{proof}

We now connect the new function $L(n,k,q)$ with $L(k,q).$

{\theorem \label{thm4} For all prime powers $q,$ and all nonnegative integers $k$ we have
$$\lim_{n \rightarrow \infty }L(n,k,q)=L(k,q).$$
More precisely, there is an integer $n_0\ge L(k,q),$ such that for all $n\ge n_0$ we have $L(n,k,q)=L(k,q).$
}

\begin{proof}
By Lemmas \ref{1} and \ref{2}, the sequence $n \mapsto L(n,k,q)$ is increasing and bounded. Hence, being integral, it converges stably to a limit which can be no other than $L(k,q).$
Let $n_0$ be such that $L(n_0,k,q)=L(k,q).$ By Lemma \ref{3}, we see that $n_0\ge L(k,q).$
\end{proof}
{\bf Remark:} The computations of the Appendix suggest that such an $n_0$ can be very large. If Proposition 2 is tight then, by Theorem \ref{thm4} $n_0 \ge \frac{q^k-1}{q-1}.$
In the special case $q=2,$ the second proof of Theorem 1 shows that $n_0=2^k-1.$\\

There is a link to Delsarte's bound (equation (1)) quoted in the Introduction.
{\prop \label{HW} For all prime powers $q,$ and all integers $n\ge k\ge 1,$ we have
$$q^k \le \sum_{i=0}^{L(n,k,q)}{n \choose i}(q-1)^i.$$
Further
$$L(k,q)\le \frac{\sum\limits_{i=0}^{L(n,k,q)}{n \choose i}(q-1)^i-1}{q-1}.$$}\vspace{-0.0cm}
\begin{proof}
The first assertion is a direct application of Equation (1) in Introduction (\cite[Th. 4.1]{D}) with $|C|=q^k,$ and $s=L(n,k,q).$ Combining this result with Proposition \ref{prop1} gives the second assertion.
\end{proof}
We give an asymptotic version of the preceding results. Let $${\mathcal L}(R)=\limsup_{n \rightarrow \infty}\frac{1}{n}\log_q(L(n,\lfloor Rn \rfloor,q)).$$

{\theorem If $C_n$ is a family of codes of rate $R$ then
$${\mathcal L}(R)\le R \lambda(q) \le H_q({\mathcal L}(R)).  $$
In particular ${\mathcal L}(R)\le t(q),$ where $t(q)$ is the unique solution in the range $(0,\frac{q-1}{q})$ of $H_q(x)=x.$ See Fig. 1.
}
\begin{proof} The first inequality follows by Lemma \ref{1}, upon observing that $$ \limsup_{n \rightarrow \infty}\frac{1}{n}(L(k,q))=R\lambda(q).$$
The second inequality comes from the second assertion of Proposition \ref{HW}, after using standard entropic estimates \cite{HP}.
The second assertion is obtained by combining the first and second inequality.
\end{proof}

Define the domain $\mathcal D$ as the set of points in the plane $(R,\mathcal L)$ that are realized by a family of codes. By the preceding result, this domain is contained in the domain of
boundaries given by, counterclockwise, in Fig. 2 by
\begin{enumerate}
\item the straight line ${\mathcal L}=R$ from $R=0$ till $R=t(q),$
\item the horizontal line ${\mathcal L}=t(q)$ from $R=t(q)$ till $R=1,$
\item  the vertical line $R=1$ from ${\mathcal L}=t(q)$ till ${\mathcal L}=\frac{q-1}{q},$
\item the curve ${\mathcal L}=H_q^{-1}(R),$ from $R=1$ till $R=0.$
\end{enumerate}
\vskip 1mm
\begin{center}
\includegraphics[width=10cm,height=4.5cm]{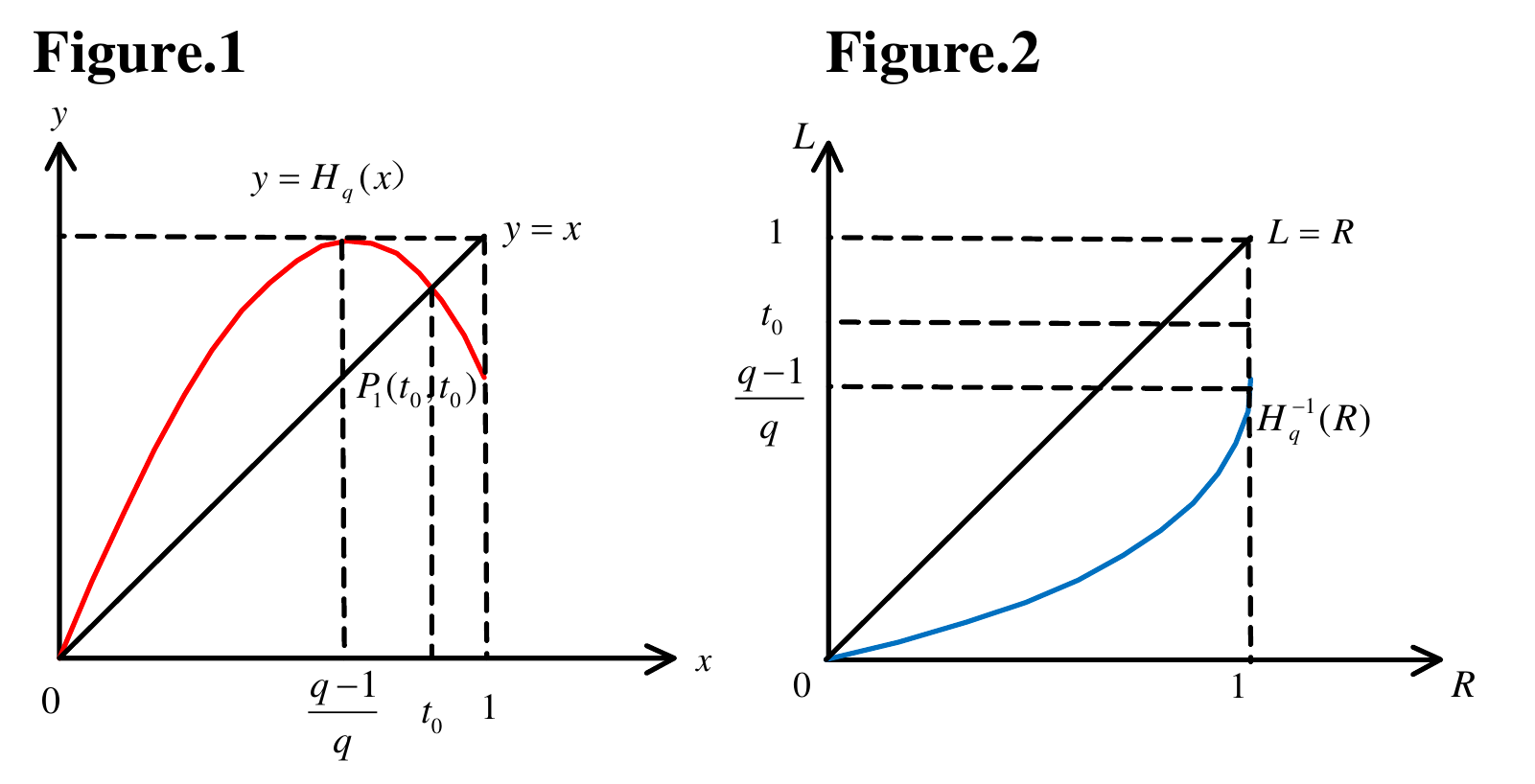}
\end{center}
\vskip 1mm
\vskip 0.0000001cm

Determining the domain $\mathcal D$ explicitly, in the same way as the domain of packing and covering codes in \cite{CHLS} is a challenging open problem.

\section{Nonlinear codes}
{\bf Warning:} In this section only $q$ is an arbitrary integer $>1.$

The nonlinear analogue of the function $L(k,q)$ is the function $N(M,q)$ which is the largest number of distances between two codewords of an unrestricted code of size $M$ over
some finite alphabet $A_q$ of size $q.$
This function is completely determined in the following Theorem.
{\theorem \label{tight2} For all integers $M\ge 2,$ we have
$$N(M,q)={M \choose 2}.$$ }\vspace{-0.8cm}
\begin{proof}
By definition we have immediately $N(M,q)\le{M \choose 2}.$ By an inductive process, we construct a code $C_M$ with ${M \choose 2}$ distances. To simplify matters take $q=2.$
We search for codes in a special form where nonzero codewords are of the form $(1,1,\dots,1,0,\dots,0),$ that is a run of ones followed by a run of zeros. Thus the distance between two such codewords is equal to the difference of their weights. For $M=2,$ we may take the length $1$ code $\{0,1\}.$
Assume $C_M$ is constructed with codewords of successive weights $w_0=0<w_1<\cdots<w_{M-1}.$ We construct a code $C_{M+1}$ by adding a tail of zeros to $C_M$ on the right, of length to be specified later, and by adding a new codeword of weight $w_M.$ The new distances are $M$ in number, given by
$w_M, w_M-w_1,\dots,w_M-w_{M-1}.$ These distances are pairwise distinct because $(w_M-w_i)-(w_M-w_j)=w_j-w_i.$ To make sure they are distinct from the distances in $C_M,$ we must check that
$$ (w_M-w_i)\neq w_j-w_k,$$ with $i,j,k$ distinct nonegative integers $\le M-1.$ This is enforced if we take $w_M$ large enough. This condition on $w_M,$ in turn, will determine how long the tail must be.
Since ${{M+1} \choose 2}-{M \choose 2}=M,$   we are done.
\end{proof}

The nonlinear analogue of the function $L(n,k,q)$ is the function $N(n,M,q)$ which is the largest number of distances between two codewords of an unrestricted code of size $M$
and length $n$ over some alphabet $A_q,$ of size $q.$

The analogue of Theorem \ref{thm4} in this context is as follows. The proof is similar and omitted.

{\theorem For all integers $q>1,$ and all nonnegative integers $M$ we have
$$\lim_{n \rightarrow \infty }N(n,M,q)=N(M,q).$$
More precisely, there is an integer $n_0\ge N(M,q),$ such that for all $n\ge n_0$ we have $N(n,M,q)=N(M,q).$
}\\

Denote by $N_0(M,q)$ the smallest integer $n$ such that $N(n,M,q)=N(M,q).$
{\prop If $M-1$ is a power of a prime, then $N_0(M,q)\le 2N(M,q)+1.$ }
\begin{proof}
 Assume $M=s+1,$ where $s$ is a power of a prime. We know there is a Singer difference set \cite{S} $S=\{v_0, v_1, \dots, v_{s+1}\},$ with parameters $(s^2+s+s,s+1,1).$
Consider the $s+1$ by $s^2+s+1$ matrix with rows $g_i,$ when $g_i$ contains $v_i$ consecutive ones to the left and zeros elsewhere. The Hamming distance from $g_i$ to $g_j$
is $|v_i-v_j|.$ The code formed by the $M$ rows of this matrix has length $s^2+s+1=M^2-M+1=2{M \choose 2}+1$ and $M \choose 2$ distances, by the design property. Hence, in this case,
$n_0\le 2{M \choose 2}+1.$ For instance, if $s=2,$ we have $S=\{1,2,4\},$ and the code is $\{1000000,1100000,1111000\}.$ See \cite[p.264]{BJL} for details on, and examples of
Singer difference sets.
\end{proof}
Denote, for any integer $t,$ by $pp(t)$ the smallest prime power $\ge t.$

{\cor \label{NT} For all integers $M>1,$ we have $$N_0(M,q)\le 2N(pp(M-1)+1,q)+1\le 2N(2M,q)\sim 8 {M \choose 2}.$$}\vspace{-1.0cm}
\begin{proof}
 We claim that $N_0(M,q)$ is a nondecreasing function of $M.$ The first inequality will follow by the previous theorem, since $M\le pp(M-1)+1.$ To prove the claim note that, if we have a set of $M+1$ vectors of length $N_0(M+1,q),$
 with ${M+1 \choose 2 }$ distances,
 removing any vector will result into a set of $M$ vectors with ${M+1 \choose 2 }-M={M \choose 2}$ distances. Hence $N_0(M,q)\le N_0(M+1,q).$ The second inequality follows
 by the crude bound $pp(x)\le 2x,$ valid for any positive integer $x.$
\end{proof}

{\bf Remark:} It is possible to reduce the upper bound on $pp(x)$ to $pp(x)\le x+x^a,$ with $a=0.525,$ building on recent estimates on the existence of primes in short intervals
\cite{BHP}. This sharpens the upper bound on $N_0(M,q)$ to $2N(M+O(M^a),q)+1\sim 2 {M \choose 2},$ for $M\rightarrow \infty.$
\section{Conclusion and open problems}
In this note, we have studied a  problem of extremal combinatorics: maximizing the number of distinct nonzero weights a linear code can have.
We conjecture, based on extensive numerical calculations on very long codes, that the bound of Proposition 2 is tight but cannot prove it. A proof was found later in \cite{proof}.
A recursive approach in the manner of the proof of Theorem \ref{tight2} would require to produce $q^k$ new weights to go from $L(k,q)$ to $L(k+1,q).$
But a code achieving $L(k,q)$ has only $\frac{q^k-1}{q-1}<q^k$ distinct weights. Thus establishing the tightness of Proposition 1 is the main open problem of this note.
Sharpening the upper bound on $N_0(M,q)$ of Corollary \ref{NT} is also a challenging question.
Determining explicitly the domain $\mathcal D$ of Section 5 seems to require better lower bounds on $L(n,kq)$ that those at our disposal.

{\bf Appendix: numerical examples}

We provide lower bounds on $L(k,q)$ by computing the number of weights in long random codes produced by the computer package Magma \cite{M}.
\begin{table}
\begin{center}
\noindent\caption{Proposition \ref{expo}}\vspace{0.2cm}
\begin{tabular}{|c|cccccccccc|}
  \hline
   $k $                         & 3       & 4      & 4      & 6       & 6       & 10        & 10        & 12     & 12      & 12 \\
  \hline $q $                   & 3       & 5      & 8      & 9       & 13      & 16        & 25        & 29     & 49      & 121  \\
  \hline $L(k,q)\ge$          & 11      & 29     & 41     & 177     & 241     & 4609      & 6913      & 31745  & 52225   & 125953  \\
  \hline
\end{tabular}
\end{center}
\end{table}
We give some numerical examples in Table 1 about the lower bound of Proposition \ref{expo}.

When $n$ is in the millions, we can find linear $[n,k]_q$-codes that meet the upper bound in Proposition \ref{prop1}: see Table 2.
\begin{table}
\begin{center}
\noindent\caption{ $n=6 \,000\, 000$ } \vspace{0.2cm}
\begin{tabular}{|c|cccccccccccc|}
  \hline
   $k $                         & 3       & 3      & 3   &3     & 3     &3  &3    & 4       & 4        & 4        & 5     & 5       \\
  \hline $q $                   & 3       & 4      & 5   &7     & 8     &9  &11    & 3      & 4        & 5        & 3     & 4        \\
  \hline $L(k,q)=$               & 13      & 21     & 31  &57    & 73    &91 &133    & 40     & 85      & 156      & 121  & 341    \\
  \hline
\end{tabular}
\end{center}
\end{table}
\end{document}